\documentclass[runningheads]{llncs}
\usepackage{graphicx}
\usepackage{amsmath}
\usepackage{amsfonts}
\usepackage{amssymb}
\usepackage{subfig}
\usepackage{hyperref}
\usepackage{algorithm}
\usepackage{algorithmic}
\usepackage{verbatim}
\begin{document}
\author{Dong Ma\inst{1} \and
Xixiang Lyu\inst{1} \and
		Renpeng Zou\inst{1}\orcidID{0000-0002-2759-1328}}
\authorrunning{F. Author et al.}
% First names are abbreviated in the running head.
% If there are more than two authors, 'et al.' is used.
%
\institute{Xidian University, Xifeng Road, Xi'an City, Shaanxi Province, China \\
	\email{xxlv@mail.xidian.edu.cn} \\
	\email{rpzou@stu.xidian.edu.cn}}
\title{A Novel Variable K-Pseudonym Scheme Applied to 5G Anonymous Access Authentication}
\titlerunning{Variable K-pseudonym Scheme in 5G Anonymous Access Authentication}
%
%\titlerunning{Abbreviated paper title}
% If the paper title is too long for the running head, you can set
% an abbreviated paper title here
%
\maketitle              % typeset the header of the contribution
\begin{abstract}
Anonymous access authentication schemes provide users with massive application services while protecting the privacy of users' identity. The identity protection schemes in 3G and 4G are not suitable for 5G anonymous access authentication due to complex computation and pseudonym asynchrony. In this paper, we consider mobile devices with limited resources in the 5G network and propose an anonymous access authentication scheme without the Public Key Infrastructure. The anonymous access authentication scheme provides users with variable shard pseudonyms to protect users' identities asynchronously. With the variable shared pseudonym, our scheme can ensure user anonymity and resist the mark attack, a novel attack aimed at the basic k-pseudonym scheme. Finally, we analyze the scheme with BAN logic analysis and verify the user anonymity. 

\end{abstract}
\section{Introduction}
With the development of mobile network technology, mobile networks~\cite{MNET} have become an indispensable part of people's life. According to \cite{cerwall2016ericsson}, there were 3.9 billion smartphones globally in 2016, which is estimated to rise to 6.8 billion by 2022. With smartphones, people can communicate with others easily and search for information quickly. However, due to the openness of wireless networks, users' identities cannot be protected effectively~\cite{sa,ahlawat2018investigating}. Once a user's International Mobile Subscriber Identification Number (IMSI) is intercepted, the adversary can track the user automatically and launch the man-in-the-middle attack (MITM) to steal the user's private information. Moreover, because of the inherent mobility, users need to be authenticated frequently but without safe identity privacy protection. 

As a result of lacking identity privacy protection, a user can be tracked by some organizations without the user's authorization. What's worse, the organizations may share the private information with other malicious parties which violate the user's privacy. For example, a user authorizes a semi-trusted mobility management entity (MME) to access his location information which is associated with his/her IMSI.
MME may share the information to third parties such as the local tourist office and the advertising agencies, which will send their advertisements to nearby users without their authorizations.
 
For user identity privacy, there are some schemes in existing literature. 1) GSM system~\cite{gsm} uses Temporary Mobile Subscriber Identity (TMSI), instead of IMSI. Because a user need to update TMSI frequently at different Visitor Location Register (VLR) with his IMSI, the adversary can intercept and capture the user's identity. 2) In 4G Long Term Evolution (LTE) Network~\cite{lte1,lte2}, Globally Unique Temporary Identity (GUTI) is adopted as the temporary identity of the user equipment (UE)~\cite{GUTI}. But the UE has to send IMSI to get or retrieve the temporary GUTI in some situations, so the UE' identity is still at risk of being revealed. 3) Public-key based schemes are not suitable for the application scenario of 5G access authentication, because they need the support of Public Key Infrastructure (PKI) and execute some complex mathematical operations such as exponent operations and bilinear pairing operations~\cite{analy}. Considering mobile devices are limited with resources, we present a shared key based anonymous access authentication scheme, which not only avoids complicated calculations, but also guarantees user anonymity.

Constrained by the existed structure of 5G access authentication and the limited capabilities of users, we propose a shared key based anonymous authentication scheme. Our contributions are as follows.
\begin{itemize}
\item [$\bullet$] We propose a 5G anonymous access authentication scheme based on shared keys. By the shared keys, UE and HSS can distinguish the valid shared pseudonym from the variable k-pseudonym sets, while the using pseudonym is still a secret for others, including MME. 
\item [$\bullet$]We present a robust anonymous access authentication scheme. Owing to the variable k-pseudonym sets, the UE can choose the suitable size of the k-pseudonym sets according to the actual network environment.
\item [$\bullet$]We design the shared pseudonym to resist the intersection attack and the mark attack. The intersection attack and the mark attack will be described in Sect.~\ref{sana}. By the shared pseudonym, the UE utilizes a dynamic temporary identity, while the adversary cannot link the variable k-pseudonym sets with the UE's identity, guaranteeing the robustness of our scheme.
\item [$\bullet$]We analyze the scheme with BAN logic analysis. After the careful derivation process, we conclude that UE and HSS can reach an agreement on the UE's identity, including the shared pseudonym.
\end{itemize}
\noindent \textbf{Organization of the Paper.} Sect.~\ref{related} reviews the related literatures. Sect.~\ref{back} provides the relevant background materials. In Sect.~\ref{vkp}, we introduce the proposed protocol in detail. In Sect.~\ref{sana}, we analyze the security and logical correctness of our scheme, respectively. At the last section, we highlight some concluding remarks. 

\section{Related Works} \label{related}
Anonymous access authentication in mobile communication networks has captured attentions of researchers and practitioners recently~\cite{ferrag2017security,alliance20155g,khan2018imsi,rupprecht2018security}. In~\cite{S3161379}, researchers introduced DHIES (Diffie-Hellman Integrated Encryption Scheme) into authentication, protecting user identity. In~\cite{PKI}, the authors introduced PKI into the EPS-AKA authentication process, which is adopted in 4G LTE Network. This change ensures user identity never being released as plaintext in untrusted networks. 

Based on KP-ABE~\cite{goyal2006attribute}, authors in~\cite{S3162108} suggest an implementation in which there is one global entity namely AuS (Authentication Server) for all operators. AuS has to generate the public key and a private key for each operator.  In~\cite{IBE}, considering that public-key based solutions have a higher cost both in terms of communication and computation, Khan et al. proposed a modified solution by using the identity based encryption (IBE). Since the public parameters have only local significance,  several public keys need to be securely provisioned to the UE, increasing user burden. 

In~\cite{PSE}, Norrman et al. presented a new scheme by establishing a series of pseudonyms between UE and HSS. The solution can reduce impact on the bandwidth compared to public-key solutions. But when lost or asynchronous pseudonyms happen, the public-key technologies have also been considered as a potential approach to solving the problem. 

To avoid the complex public-key calculations, Li et al. proposed an anonymous authentication scheme based on a shared key in~\cite{kli}. Besides, the authors presented the enhanced Dolev-Yao model and introduced the intersection attack. By the static k-pseudonym set, they fixed the intersection attack basically. Inspired by this scheme, we try to design the shared pseudonym to construct variable k-pseudonym sets, resisting the intersection attack.

\begin{comment}
In~\cite{2pki}, the Diffie-Hellman (DH) key agreement process was added between the UE and the MME. Thus the UE can encrypt its identity with the negotiated key, protecting the UE's identity.

Based on the public-key encryption, \cite{PKI} protects the privacy of UE's identity with high computational overhead, causing the HSS to suffer the DoS (Denial of Service) attack.

is a public-key scheme in which a single public key can be associated with several private keys, the attributes which match the access structure will be able to decrypt.

IBE can be seen as a form of attribute-based encryption in which the attributes used in the system is directly the identity of the recipient.

 which can both accomplish the identification and mutual authentication
 
 as a result of the UE's mobility,
 
 Using the k-pseudonym set, they proposed an authentication protocol with comprehensive consideration of the limited storage and computation capacity of mobile devices. 
\end{comment}
 
\section{Background} \label{back}
In this section, we introduce some preliminaries including the basic k-pseudonym scheme and ZUC algorithms. The symbols used in the paper are shown in Table~\ref{symbol}. 
\begin{table}[htbp] 
	\caption{Notation Summary} \label{symbol}
	\begin{tabular}{c|l}
		\hline
		\multicolumn{1}{l}{Symbol} & Description                                            \\
		\hline
		UE         & User Equipment                                                         \\
		HSS        & Home Subscription Server                                               \\
		MME        & Mobility Management Entity                                             \\
		Key        & the shared key between a UE and the HSS                                      \\
		IMSI       & International Mobile Subscriber Identification Number \\                 		                                 
		H()        & a collision resistant hash function                                                          \\
		$HMAC$     & a collision resistant hash function with a cipher key                                        \\
		${\scriptsize HMAC_{40}}$& a collision resistant hash function with a cipher key, select the high 40 bits of the output\\ 
		$H_i$      & $HMAC(Key||P_i)$                                              \\ 
		$f_i()$    & the subfunctions of Milenage algorithm which is used in the authentication of 5G network \\
		$P_0$      & the anchor shared pseudonym                                            \\   
		$P_i$      & the shared pseudonym used in the i-th authentication                       \\ 
		$\{P_i\}$  & a k-pseudonym set including $P_i$                                      \\ 
		$SQN_{IMSI}$    & the SQN of IMSI                 \\   
		$SQN_0$    & the SQN of $P_0$                 \\       
		$count_i$  & count the number of ZUC has been run in the i-th authentication        \\                
		\hline
	\end{tabular}
\end{table}
\begin{figure}[h]
\centering
\includegraphics[width=0.9\linewidth]{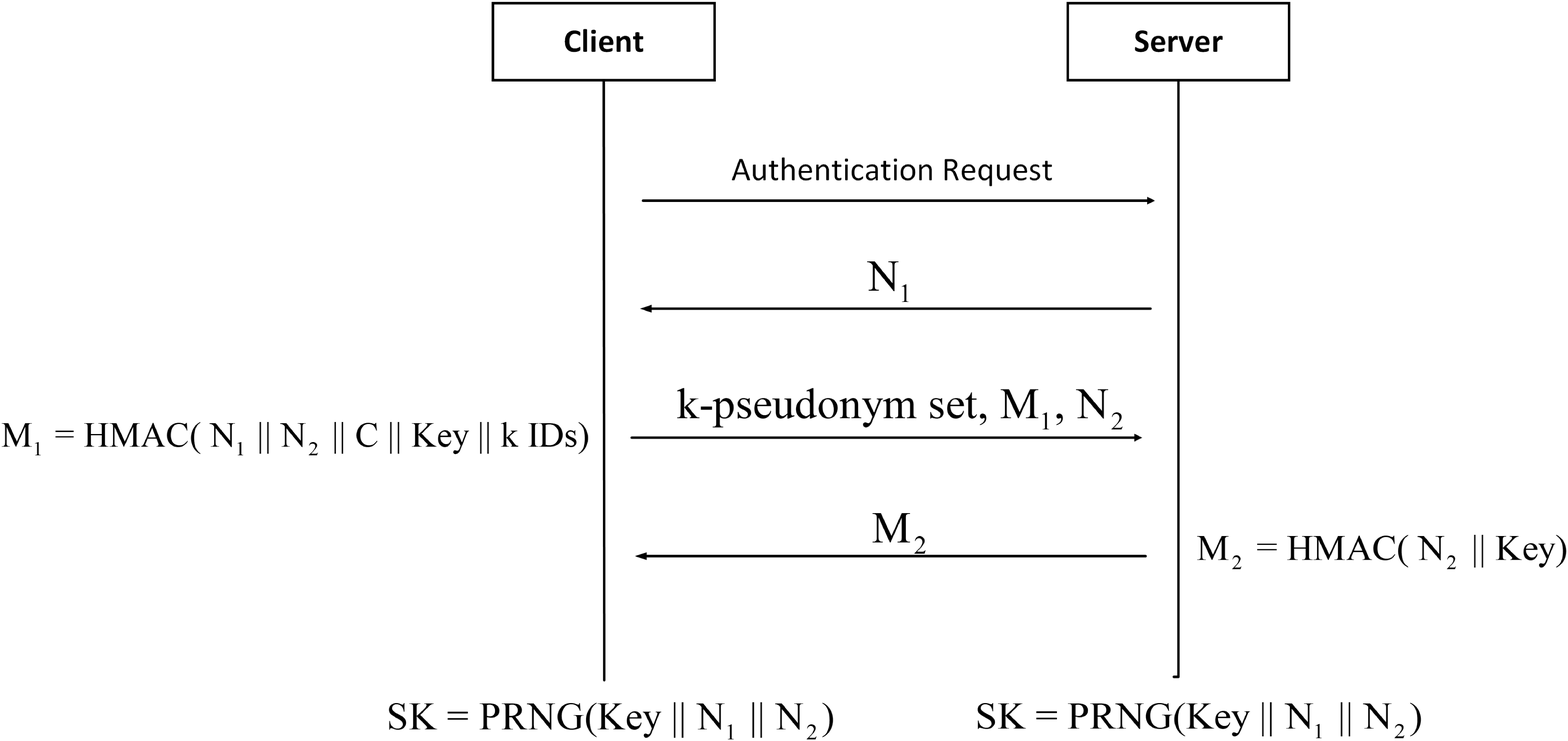}
\caption{The anonymous authentication process using the k-pseudonym set}
\label{fig:basick}
\end{figure}

\subsection{K-pseudonym Scheme}
In the k-pseudonym scheme, a user sends the k-pseudonym set including his identity and the message encrypted by the shared key, a key obtained from server securely. The server traversals the shared keys according the k identities in the identity set one by one and verify the authentication information respectively by the corresponding keys. Once the authentication information is verified correctly, the user is authenticated. Fig.~\ref{fig:basick} shows the authentication process of the k-pseudonym scheme.

\noindent 1) user $\rightarrow$ server: a user sends an authentication request to the server.

\noindent 2) server $\rightarrow$ user: upon receiving the authentication request, the server generates a random number $N_1$ and sends it to the user as a challenge. 

\noindent 3) user $\rightarrow$ server: on receiving the random number $N_1$, the user generates a random number $N_2$, and calculates $M_1$ by Eq.~(\ref{M1}).
\begin{equation}
M_1=HMAC(N_1||N_2||C||Key||(k-pseudonym \ set)) \label{M1}
\end{equation}
Then the user sends the k-pseudonym set including his real identity and other k-1 assistant pseudonyms, the random number $N_2$, and $M_1$ to the authentication server.

\noindent 4) server $\rightarrow$ user: receiving the message from the user, the server calculates the corresponding $M_1'$ in Eq.~(\ref{M11}), where $each \ ID$ is an identity in the k-pseudonym set, $Key_{ID}$ is the shared key related with this identity. Finally, the server verifies whether $M_1'$ is equal to $M_1$.
\begin{equation}
M_1'=HMAC(N_1||N_2||each \ ID||Key_{ID}||(k-pseudonym \ set))  \label{M11}
\end{equation}
\noindent If $M_1'$ is equal to $M_1$, the server can determine that the corresponding ID is the user's real identity and complete the authentication process. After that, the server calculates $M_2$ in Eq.~(\ref{m2}).
\begin{equation}
M_2=HMAC(N_2||Key) \label{m2}
\end{equation}
In the end, the user calculates $M_2'=HMAC(N_2||Key)$ according to the random number $N_2$ and the shared key. And then he verifies whether $M_2'$ is equal to $M_2$ received from the authentication server. If the verification is successful, the user and the server complete the mutual authentication and generate the session key $SK=PRNG(Key\oplus N_1\oplus N_2)$, where $\oplus$ represents a xor operation.

\subsection{ZUC Algorithm}
As a stream cipher algorithm, the ZUC algorithm has been adopted as the kernel of the third set of the LTE cryptographic algorithms \cite{zuc1}. It consists of three layers and initializes the internal states by a 128-bit cipher key $K$ and a 128-bit initialization vector $IV$. In this paper, we use ZUC to generate the variable shared pseudonyms. Here we briefly introduce the process of ZUC.
\begin{itemize}
\item[$\bullet$] Linear feedback shift registe (LFSR) is constructed from 16 register units, each holding 31 bits. And the feedback is defined by a primitive polynomial over the finite field GF($2^{31}-1$).
\item[$\bullet$]  Bit reorganization(BR) extracts 128 bits from the states of the LFSR and forms four 32-bit words, where the first three words will be used by the nonlinear function F in the bottom layer, and the last word will be involved in producing the key stream. It forms 4 of 32-bit words $X_0$ , $X_1$ , $X_2$ , $X_3$, from the following 8 LFSR registers $s_0$ , $s_2$ , $s_5$ , $s_7$ , $s_9$ , $s_{11}$ , $s_{14}$ , $s_{15}$.
\item[$\bullet$]  Nonlinear function(F) is based on two 32-bit registers R1 and R2. The operation of F involves input from BR and uses two S-boxes $S_0$ and $S_1$. The mixing operations are the exclusive OR, the cyclic shift and the addition modulo $2^{32}$ (which takes the symbol $\boxplus$ as the modulo $2^{32}$ addition). By W=( $X_0$ $\oplus$ $R_1$ ) $\boxplus$  $ R_2$, we get the keystream word Z as $Z=W \oplus X_3$.
\end{itemize}
\begin{comment}
 Based on a shared key, 
 
 The k-pseudonym scheme was presented by Li et al. in~\cite{kli}. In the implement of the scheme, the authors introduced the enhanced Dolev-Yao model, including the intersection attack which is analyzed in Sect.~\ref{iner_attack}.
 
 Under this premise, it is not necessary for the user to be anonymous to the authentication server.
 
 Moreover, we emphasize that the attacker can participate in the system as a legitimate user or an unauthorized user.
 
 and Table~\ref{ksymbol} shows the symbols used in the k-pesudonym scheme.
 \begin{table}[h]
 \caption{Symbols explanation for k-pesudonym proposal} \label{ksymbol}
 \centering
 \begin{tabular}{c|l}
 \hline
 \multicolumn{1}{l}{Symbol} & Description                                            \\
 \hline
 $N_1$         & random number generated by the server                       \\ 
 $N_2$         & random number generated by the user                       \\ 
 C             & the user's actual identity                                   \\
 Key        & the shared key                                     \\
 SK       & the session key                 \\
 HMAC()        & a message authentication code function \\
 PRNG()        & a pseudorandom function \\     
 \hline
 \end{tabular}
 \end{table}
 
 These security algorithms are used to ensure the data confidentiality and integrity during message transmission over the air interface in the LTE Network. 
 
 The most important characteristic of ZUC is that it can output a 32-bit keystream continually and independently. 
\end{comment}
 
\section{The Proposed Scheme} \label{vkp}
In this section, we introduce the variable k-pseudonym scheme in detail. Firstly, we design the shared pseudonyms with the help of ZUC. Next, we briefly introduce the variable k-pseudonym set construction. Then the detailed process of variable k-pseudonym scheme is described. Finally we adopt the anchor shared pseudonym $P_0$ in the recovery mechanism. 

\subsection{Shared Pseudonym} \label{shared}
In the scheme, we use the predicable property of the pseudo-random sequence. We assume that UE and HSS can initialize ZUC with same shared information: the shared key $Key$ and the sequence number $SQN$. Then they can get the same pseudo-random sequence and generate the shared pseudonym synchronously but independently. Here follows the generation of the shared pseudonym.

\noindent 1) Generate the initial parameters. ZUC is initialized by a 128-bit cipher key $K$ and a 128-bit initialization vector $IV$. In the scheme, we take the shared key as $K$, and get $IV$ by the Eq.~(\ref{IV}).
\begin{equation}
\begin{split}
& Rand = H(SQN_{IMSI}) \ or \ H(SQN_{P_0})\\
& CK_0 = f_3(Key,Rand)\\
& IK_0 = f_4(Key,Rand)\\
& IV =CK_{0H}||IK_{0L} \label{IV}
\end{split}
\end{equation}

\noindent where $f_3$, $f_4$ is the subfunctions of Milenage algorithm~\cite{3gpp3g,kim2013design}, which is used in the authentication of 5G network, $CK_{0H}$ means the high 64 bits of $CK_0$, $IK_{0L}$ means the low 64 bits of $IK_0$.

\begin{figure}
	\centering
	\includegraphics[width=0.6\linewidth]{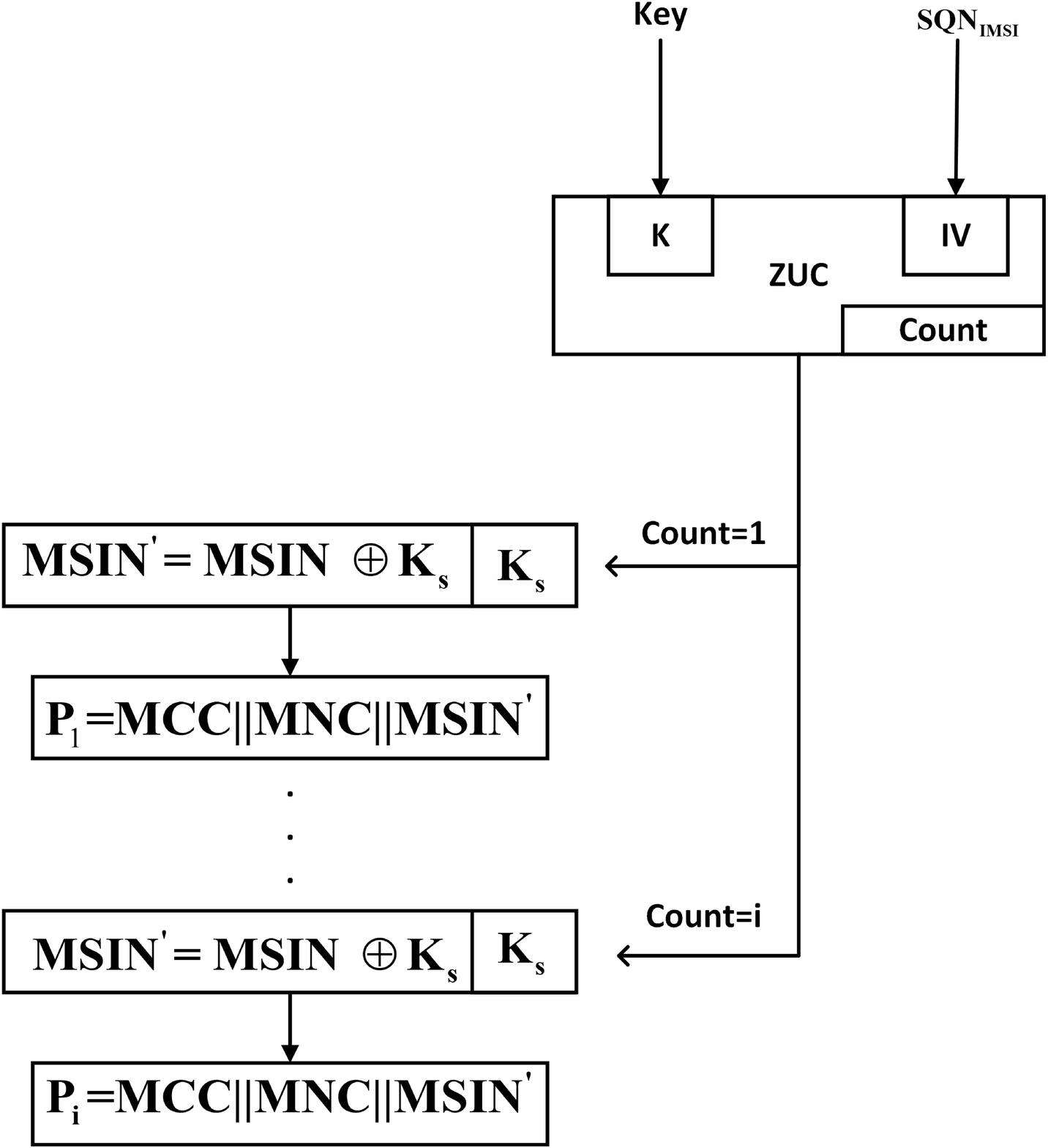}
	\caption[000]{The basic structure of updating the shared pseudonym}
	\label{fig:pse}
\end{figure}

\noindent 2) Update the shared pseudonym. As defined by 3GPP, IMSI is composed of three components: Mobile Country Code (MCC), Mobile Network Code (MNC), Mobile Subscriber Identification Number (MSIN). It is not necessary to change MCC and MNC, so we update the shared pseudonym by encrypting the MSIN of IMSI. Another thing we emphasize is that MSIN is 40 bits, so we need expand the 32-bit keystream output from ZUC to 40 bits. Thus we can encrypt the MSIN with the 40-bits expanded-keystream ($K_s$). Here we use XOR as the encrypt algorithm and $MSIN'$ is calculated by Eq.~(\ref{MSIN}).
\begin{equation}
MSIN' = MSIN \oplus K_s \label{MSIN}
\end{equation}
UE and HSS get the shared pseudonym $P_i$ by Eq.~(\ref{pi0}), where MCC and MNC  are obtained from IMSI.
\begin{equation}
P_i = MCC || MNC|| MSIN' \label{pi0}
\end{equation}
When a UE accomplish authentication with the shared pseudonym $P_i$, the UE and the HSS update the shared pseudonym $P_i$ to get the next shared pseudonym $P_{i+1}$. Fig.~\ref{fig:pse} shows the basic structure of updating the shared pseudonym. 
\subsection{Variable K-pseudonym Set Construction}
For simplicity, it is a rational assumption that a UE can get enough available assistant identities from the HSS. If the UE has connected with the HSS, the HSS sends the updating shared pseudonyms which are used by others to the UE. Considering the situation that the UE is new for the HSS, the UE should generate the k-pseudonym sets by itself. Since the UE can generate the shared pseudonym by ZUC, similarly, he can generate assistant identities.

\subsection{Authentication with Variable K-pseudonym Scheme} \label{access_with_k}
\begin{figure}
\centering
\includegraphics[width=1.0\linewidth]{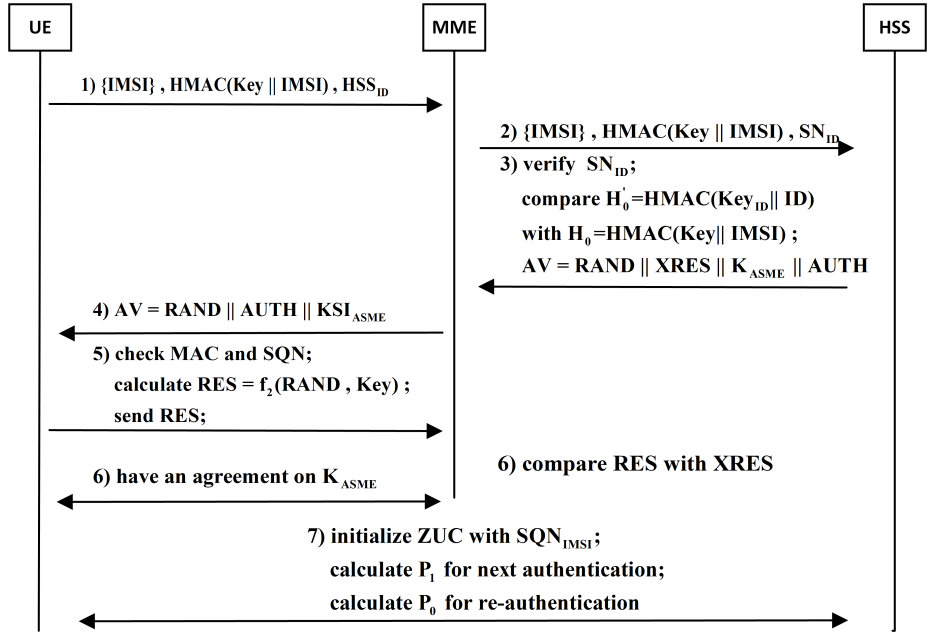}
\caption[000]{The authentication process when UE is new for HSS}
\label{fig:liucheng}
\end{figure}

Here we briefly introduce the anonymous access authentication process with the variable k-pseudonym scheme. Fig.~\ref{fig:liucheng} shows the situation when a UE is new for the HSS.

\noindent 1) UE $\rightarrow$ MME: when the UE is new for the HSS, the UE generates assistant identities to construct \{IMSI\}. Then the UE sends the k-pseudonym set \{IMSI\}, $H_0=HMAC(Key||IMSI)$, the identifier of the HSS $HSS_{ID}$ to a MME. 

\noindent 2) MME $\rightarrow$ HSS: the MME forwards \{IMSI\}, $H_0$ to the target HSS and replaces $HSS_{ID}$ with its own $SN_{ID}$.

\noindent 3) HSS $\rightarrow$ MME: when receiving the authentication request, the HSS checks the $SN_{ID}$ to confirm the MME. Then the HSS traverses those identities included in \{IMSI\} to find the UE's IMSI by compared $H_0'=HMAC(Key_{ID}\-||ID)$ with $H_0$, where ID is the identity in the \{IMSI\} and $Key_{ID}$ is the key bounded with the ID. If the HSS cannot find the  $Key_{ID}$, then the HSS ignores the ID. Once finding an ID that makes $H_0' = H_0$ and is in \{IMSI\}, the HSS authenticate the UE. After that, the HSS calculates a authentication vector $AV$~\cite{av} by Eq.~(\ref{av}), helping the MME accomplish the final authentication. Finally, the HSS sends the $AV$ to the MME. Here we define the $ SQN $ used in the first access authentication as $SQN_{IMSI}$, which participates in the update of shared pseudonyms.
\begin{equation}
\begin{split}
& AUTH = (SQN \oplus AK) || AMF || MAC \\
& AV = Rand || XRES || K_{ASME} ||AUTH \label{av}
\end{split}
\end{equation}
\noindent 4) MME $\rightarrow$ UE: upon receiving the authentication response, the MME gets $ Rand $, $ AUTH $ and $K_{ASME}$ from the $AV$. Then the MME assigns a 3 bits key identification ($KSI_{ASME}$) for $K_{ASME}$ and sends $Rand || AUTH || KSI_{ASME}$ to the UE.

\noindent 5) UE $\rightarrow$ MME: when receives the authentication response, the UE checks $MAC$ and $SQN$ in the $ AUTH $. If they are matched, the UE calculates $RES$ with $f_2(Rand,Key)$, where $f_2$ is a secure function shared with the HSS. Finally the UE sends $RES$ to the MME. Because only the UE and the HSS can calculate $AK$ and get the $SQN$, so they can initialize ZUC with the $SQN_{IMSI}$ privately.

\noindent 6) MME $\rightarrow$ UE: the MME compares $RES$ with $XRES$. If $RES = XRES$, the MME sends an authentication complete signal to the UE. After the authentication finished, the UE and the MME have an agreement on $K_{ASME}$, building a secure link between the UE and the MME. 

\begin{figure}
\centering
\includegraphics[width=0.8\linewidth]{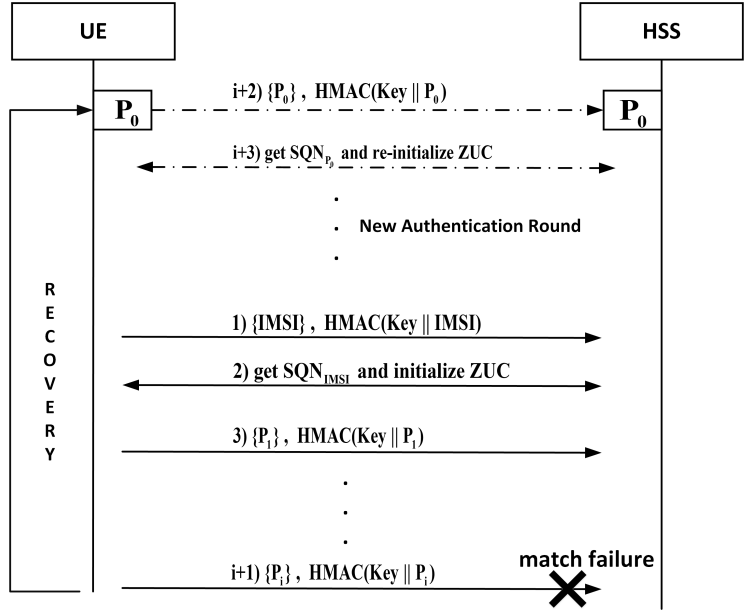}
\caption[000]{The usage of shared pseudonyms}
\label{fig:pseudonym}
\end{figure}

\noindent 7) UE $\leftrightarrow$ HSS: if $H_0' \neq H_0$ or $RES \neq XRES$, the authentication is interrupted. We assume that when the UE tries \{IMSI\} again, the UE must use the same k-pseudonym set \{IMSI\}, which means the UE has to store the whole \{IMSI\} before the IMSI is authenticated. After the UE's IMSI is authenticated, the UE and the HSS get the shared pseudonym $P_i$ ($i \geq 1$) synchronously. In the next access authentications, the UE can use the shared pseudonym $P_i$ as his temporary identity, Fig.~\ref{fig:pseudonym} briefly introduces the usage of the shared pseudonyms.
\subsection{Recovery Mechanism} \label {recovery}
When the UE tries to connect with the target HSS, the authentication failure comes from two situations: 1) there are something wrong with the UE, such as miscalculation or signal distortion. The faults lead to the shared pseudonym $P_i$ is not match with the shared pseudonym $P_i'$ generated by the HSS. 2) the HSS loses $P_i'$, because of an unexpected cleanup of memory. In order to continue the UE's access authentication process, we add the anchor shared pseudonym $P_0$, generated synchronously both at the UE and the HSS. The first we emphasize is that the values of $SQN_{IMSI}$ ($SQN_{P_0}$) is protected by the USIM’s physical security features, which means the UE and the HSS can regard $SQN_{IMSI}$ ($SQN_{P_0}$) as stable shared information~\cite{sqn}. According to this fact, we define the $P_0$ by Eq.~(\ref{p0}).
\begin{equation}
P_0 = MCC || MNC|| (MSIN \oplus HMAC_{40}(Key||SQN_{IMSI})) \label{p0}
\end{equation}

Besides, the time delay between two continuous access authentication is longer than the time of updating the shared pseudonym, because the UE only needs to be authenticated when he is back online after rebooting device or turning off flight mode. So it is not necessary to consider the time delay caused by the generation of shared pseudonym between the UE and the HSS. In view of this situation, we assume access authentication failure just comes from situation (1) and (2).

After the UE's IMSI is authenticated, the UE and the HSS calculate $P_0$ with $SQN_{IMSI}$. In other re-authentication situations, the UE and the HSS calculate $P_0$ with $SQN_{P_0}$. Once the authentication failure coming, we continue the access authentication by returning to the anchor shared pseudonym $P_0$. Moreover, $SQN_{P_0}$ ensures $P_0$ is variable at different authentication rounds, which means we can use a new \{$P_0$\} to restart the access authentication, no need to store the old \{$P_0$\}.

\begin{comment}
First, a pseudo-random sequence is a definitive sequence with random property and usually generated by the linear feedback shift register (LFSR). As a theoretical pseudo-random sequence, it is impossible to distinguish a pseudo-random sequence and a real random sequence.

given the same initial input, the generator can export the same definitive sequence with random property. Next, 

Then the UE utilize the assistant identities to construct the k-pseudonym sets including the UE's shared pseudonym.

Meanwhile the HSS tries finding the key bounded with the ID ($Key_{ID}$). 

In the access authentication process, the $SQN$ is maintained by both the UE and the HSS, 

when the UE's IMSI is authenticated, 

We assume that the UE's IMSI must can be authenticated at the HSS. When the UE tries \{IMSI\} again. So the UE  will use the same k-pseudonym set \{IMSI\}, which means the UE has to store the whole \{IMSI\} before the UE's IMSI is authenticated. 
\end{comment}
 
\section{SECURITY ANALYSIS} \label{sana}
In this section, we first analyze the intersection attack and the mark attack. Then we verify the logical correctness of the scheme with BAN. Finally, we discuss user anonymity. After the complete security analysis, we conclude that the proposed scheme can resist the intersection attack and the mark attack, while guaranteeing good user anonymity.
\subsection{The Intersection Attack} \label{iner_attack}
\begin{figure}[h]
	\centering
	\subfloat[]{\label{Fig:inter}%%
		\includegraphics[width=0.8\linewidth]{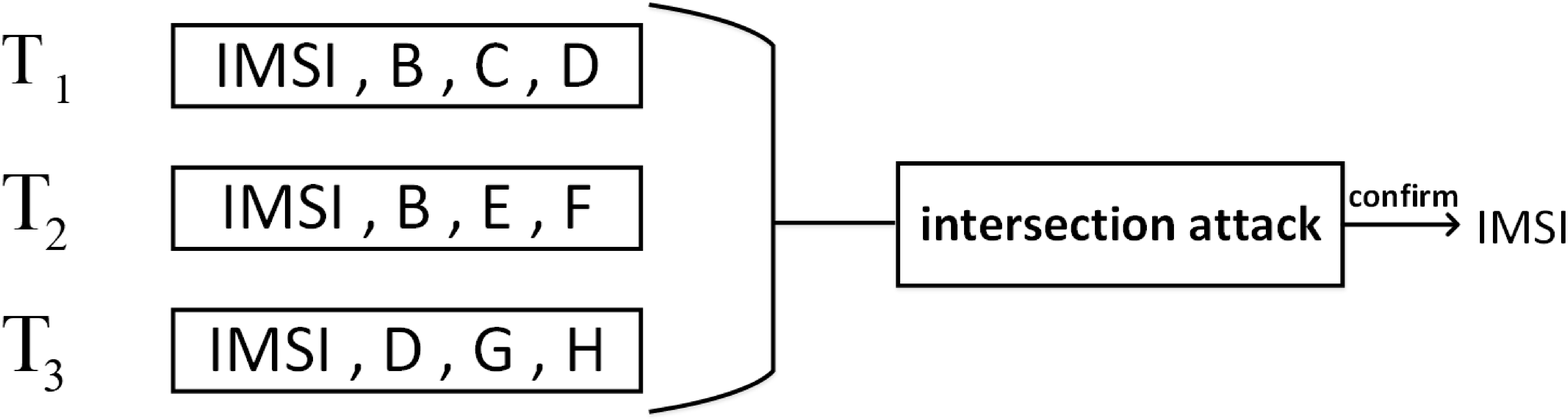}}
	\quad
	\subfloat[]{\label{Fig:avoid_inter}%%
		\includegraphics[width=0.8\linewidth]{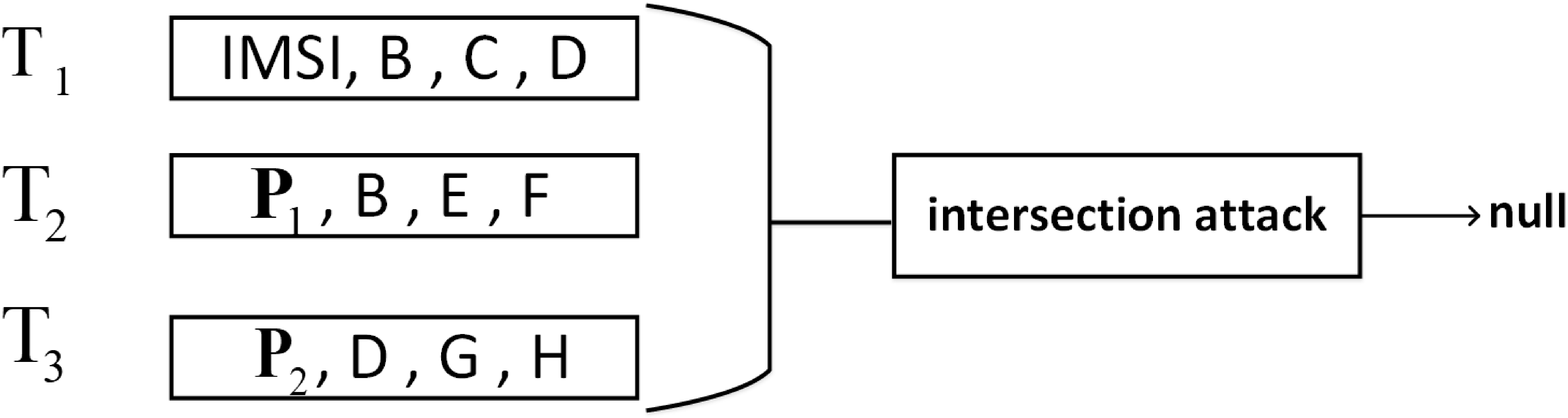}}\\
	
	\caption{\ref{Fig:inter} shows the the intersection attack; \ref{Fig:avoid_inter} avoid the intersection attack with anonymous set}
	\label{Fig:int}
\end{figure}

As considered in \cite{kli}, the intersection attack shows a situation that an adversary can observe the k-pseudonym sets generated by the target UE, and associate those relevant k-pseudonym sets with the UE's IMSI. If the UE changes the anonymous sets \{IMSI\} at different time, the adversary can reduce the range of the IMSI or even confirm it. As is shown in Fig.~\ref{Fig:inter}, 
the UE uses a k-pseudonym set \{IMSI, B, C, D\} at time $T_1$, where IMSI represents the UE's real identity and others for the assitant identities in the k-pseudonym set. Next, if the UE uses a k-pseudonym set \{IMSI, B, E, F\} at time $T_2$, the adversary links the two k-pseudonym sets to the UE and find the common elements of the two sets. After analysis, the adversary can conclude that the real identity is included in \{IMSI, B\}. What’s more, if the UE uses a k-pseudonym set \{IMSI, D, G, H\} at time $T_3$, the adversary can even get the UE's IMSI with sufficient information. Under this assumption, if the adversary can get more k-pseudonym sets from the target UE, he will have higher possibility to get the UE's IMSI.

In order to resist the intersection attack, Li et al. presented a static construction of the k-pseudonym sets in~\cite{kli}. By this way, the UE employs the same k-pseudonym set during continuous anonymous access authentications. Although this method works on the enhanced Dolev-Yao model basically, there still has some questions worthy of consideration. The most import question is the robustness of the scheme. It is known that the Quality of Service (QoS) of UE is inversely proportional to the size of pseudonym set, because the larger set results more latency which downgrades QoS. If we select a large set for an unsafe environment, it will restrict the QoS in some relatively safe envirments. But if we use a small set for an relative safe environment, it cannot guarantee the UE anonymity in a critical environment. So we conclude that the static construction of the k-pseudonym sets restricts the robustness of the scheme. 

In the proposed scheme, we resist the intersection attack by the variable k-pseudonym sets. Taking into account the application scenarios of 5G anonymous access authentication, we adopt ZUC to generate the shared pseudonyms during next authentications. As is shown in Figure.\ref{Fig:avoid_inter}, the UE uses a k-pseudonym set \{IMSI, B, C, D\} at time $T_1$, and uses a k-pseudonym set \{$P_1$, B, E, F\} at time $T_2$, where $P_1$ represents the shared pseudonym of the UE. The adversary associates the two k-pseudonym sets, but he cannot get effective information about the UE's IMSI, because the shared pseudonym changes in next authentications. By the shared pseudonym, the UE can choose the suitable size of k-pseudonym set and construct the variable k-pseudonym sets, improving the robustness of our scheme.
\subsection{The Mark Attack} \label{mark_attack}
After the further study on the basic k-pseudonym scheme, we present a novel mark attack. In~\cite{kli}, to reduce user burden, the authors suggest that the HSS generates assistant identities and sends them to the UE. Moreover, in the enhanced Dolev-Yao model, the adversary can participate in the protocol as a legitimate user, which means he can mark his identity and distinguish it from a k-pseudonym set. Under this attack condition, the HSS cannot get rid of the marked assistant identities, while the UE also cannot discriminate between the normal assistant identities and the marked assistant identities. Once the UE constructs a k-pseudonym set with marked assistant identities, the adversary has a probability greater than $ \frac{1}{k} $ to get the IMSI. Here we assume that the adversary can mark a great deal of assistant identities, but not all assistant identities. 
\begin{figure}[h]
	\centering
	\subfloat[]{\label{Fig:mark}
		\includegraphics[width=0.8\linewidth]{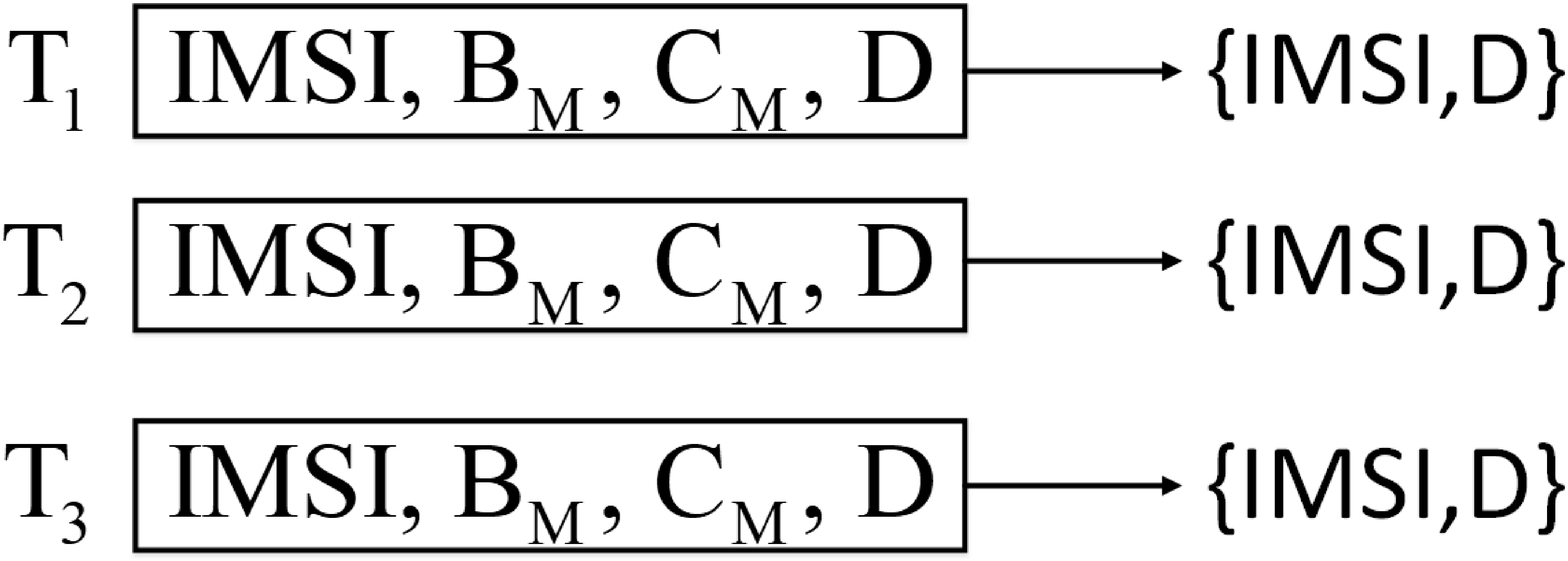}}
	\quad
	\subfloat[]{\label{Fig:avoid_mark}
		\includegraphics[width=0.8\linewidth]{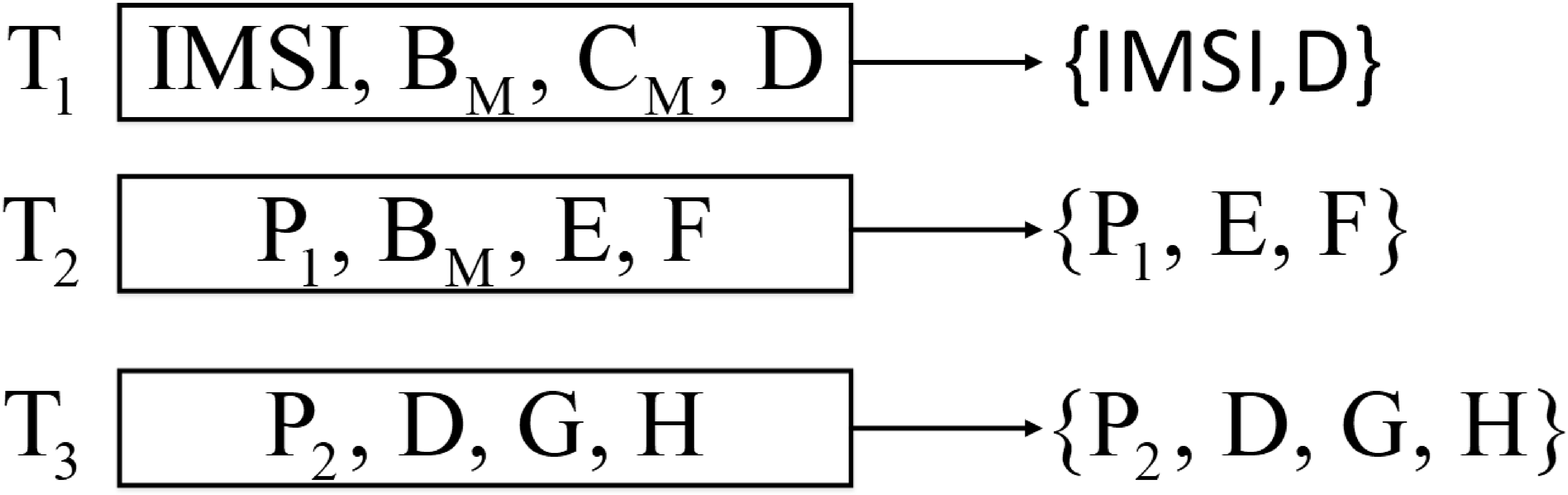}}\\
	
	\caption{\ref{Fig:mark} shows the the mark attack; \ref{Fig:avoid_mark} avoid the mark attack with variable k-pseudonym set}
	\label{Fig:int}
\end{figure}
For example, the HSS sends 100 assistant identities to the UE, and 20 of identities are marked by the adversary, including B and C. As shown in Fig.~\ref{Fig:mark}, the UE uses a k-pseudonym set \{IMSI, B, C, D\} at time $T_1$ and B, C is marked, the adversary concludes that UE's identity is in \{ID, D\}, a probability greater than $ \frac{1}{4} $. What's worse, when B, C and D are all marked, the attacker can confirm the ID directly. The adversary marks more assistant identities, he has more possibility to get the UE's identity.

In our proposal, the variable shared pseudonym is adopted to resist the mark attack. As illustrated in Figure.\ref{Fig:avoid_mark}, the UE constructs a k-pseudonym set \{IMSI, B, C, D\} at time $T_1$ and when B, C are marked, the adversary can get \{IMSI, D\}. And the UE uses a k-pseudonym set \{$P_1$, B, E, F\} at time $T_2$, only B is marked. Then the attacker only conclude that UE's identity is in \{$P_1$, E, F\}. In the best case, when the UE uses a k-pseudonym set \{$P_2$, D, G, H\} at time $T_3$, without marked assistant identities, the adversary even cannot get the \{$P_2$\}. At $T_1$, although our scheme has the same security as basic k-pseudonym scheme, \{IMSI\} only appears once in our system. In the subsequent authentications, \{$P_i$\} can use different assistant identities. After the analysis above, the UE's IMSI is hidden by the variable shared pseudonym, so our scheme can resist the mark attack.
\subsection{BAN Logic Analysis~\cite{BAN1}}
\subsubsection{BAN logical notation}
\ 

\noindent BAN logical notation used in the paper as follows:

\noindent 1) P, Q: the communication subject;

\noindent 2) X, Y: the statement or message;

\noindent 3) K: the cipher key;

\noindent 4) P $\mid \equiv$ X: P believes X; 

\noindent 5) P $\vartriangleleft$ X: P sees X;
 
\noindent 6) P $\mid \sim$ X: P said X; 

\noindent 7) P $\mid \Rightarrow$ X: P controls X;   

\noindent 8) \#(X): X is fresh;

\noindent 9) P $\stackrel{K}{\longleftrightarrow}$ Q: K is the key shared by P and Q; 

\noindent 10) $[X]_K$: the ciphertext of X encrypted by the key K. 
\subsubsection{BAN logical postulates} 
\

\noindent 1) Message-meaning rule: 
\begin{equation}
\begin{split}
\frac{P\mid \equiv Q \stackrel{K}{\longleftrightarrow} P, P\vartriangleleft [X]_K} {P\mid \equiv Q \mid \sim X} \label{mmr}
\end{split}
\end{equation}                                        
\noindent 2) Nonce-verification rule:
\begin{equation}
\begin{split}
\frac{P\mid \equiv \#(X), P\mid \equiv Q \mid \sim X} {P\mid \equiv Q \mid \equiv X} \label{nvr}
\end{split}
\end{equation} 
\noindent 3) Freshness rule:
\begin{equation}
\begin{split}
\frac{P\mid \equiv \#(X)} {P\mid \equiv \#(X,Y)}   \label{fr}
\end{split}
\end{equation} 
\noindent 4) Befief rule:
\begin{equation}
\begin{split}
\frac{P \mid \equiv Q \mid \equiv (X,Y)} {P\mid \equiv Q \mid \equiv X}   \label{br}
\end{split}
\end{equation} 
\noindent 5) Session key rule:
\begin{equation}
\begin{split}
\frac{P \mid \equiv \# K, P \mid \equiv Q \mid \equiv X} {P \mid \equiv P \stackrel{K}{\longleftrightarrow} Q} \label{sr}
\end{split}
\end{equation} 
where X here is a necessary element of K.

\noindent 6) Jurisdiction rule:
\begin{equation}
\begin{split}
\frac{P \mid \equiv Q \mid \Rightarrow X , P \mid \equiv Q \mid \equiv X} {P\mid \equiv X}   \label{jr}
\end{split}
\end{equation} 
\subsubsection{Protocol Analysis} 
\

\noindent First, the protocol can be idealized as follows:

\noindent Premise P1: HSS $\mid \equiv$ HSS $\stackrel{Key}{\longleftrightarrow}$ UE

\noindent Premise P2: UE $\mid \equiv$ HSS $\stackrel{Key}{\longleftrightarrow}$ UE

\noindent Premise P3: HSS $\mid \equiv$ UE $\mid \Rightarrow$ IMSI

\noindent Premise P4: HSS $\mid \equiv$ UE $\mid \Rightarrow P_i (i \geq 1)$

\noindent Premise P5: HSS $\mid \equiv$  $SQN_{IMSI}$

\noindent Premise P6: HSS $\mid \equiv$  $SQN_{P_0}$

\noindent Premise P5: UE $\mid \equiv$ \# $SQN_{IMSI}$

\noindent Premise P6: UE $\mid \equiv$ \# $SQN_{P_0}$

\noindent Premise P7: UE $\mid \equiv$ HSS $\mid \Rightarrow SQN_{IMSI}$

\noindent Premise P7: UE $\mid \equiv$ HSS $\mid \Rightarrow SQN_{P_0}$

\noindent The protocol flows of our scheme:

\noindent 1) UE $\mapsto$ HSS: \{IMSI\}, UE $\stackrel{Key}{\longleftrightarrow}$ HSS, $[IMSI]_{Key}$;

\noindent 2) HSS $\mapsto$ UE: $SQN_{IMSI}\oplus AK$;

\noindent 3) UE $\mapsto$ HSS: RES;

\noindent 4) (the next authentication) UE $\mapsto$ HSS: \{$P_i$\}, UE $\stackrel{Key}{\longleftrightarrow}$ HSS, $[P_i]_{Key}$;

\noindent Next, our security goals are:

\noindent $\bullet$ UE $\mid \equiv SQN_{IMSI}$

\noindent $\bullet$ HSS $\mid \equiv P_i$

\noindent $\bullet$ UE $\mid \equiv$ UE $ \stackrel{SQN_{IMSI}}{\longleftrightarrow} $HSS

\noindent $\bullet$ HSS $\mid \equiv$ UE $ \stackrel{SQN_{IMSI}}{\longleftrightarrow} $HSS

\noindent Then, analyse our scheme:

\noindent 1) Since the message-meaning rule in Eq.~(\ref{mmr}), we get:
\begin{equation}
\begin{split}
\frac{HSS \mid \equiv UE \stackrel{Key}{\longleftrightarrow} HSS, HSS \vartriangleleft [IMSI]_{Key}} {HSS \mid \equiv UE \mid \sim IMSI} \label{mo}
\end{split}
\end{equation} 
\noindent 2) After the HSS authenticates the UE's IMSI, the HSS sends an $ AV $ including $SQN_{IMSI} \oplus AK$, where $ AK $ can be calculated by Eq.~(\ref{f5}).
 \begin{equation}
\begin{split}
AK = f_5( Rand, Key )   \label{f5}
\end{split}
\end{equation}
Although $ Rand $ is transmitted as plaintext, $ Key $ is a private part, which means only the UE and the HSS can share the $SQN_{IMSI}$. Here we regard $SQN_{IMSI}$ is encrypted by the $ Key $, and according the message-meaning rule in Eq.~(\ref{mmr}), the nonce-verification rule in Eq.~(\ref{nvr}), the belief rule in Eq.~(\ref{br}) and the jurisdiction rule in Eq.~(\ref{jr}), we get:
\begin{equation}
\begin{split}
\displaystyle &\frac{UE \mid \equiv HSS \stackrel{Key}{\longleftrightarrow} UE, UE \vartriangleleft [SQN_{IMSI}]_{Key}} {UE \mid \equiv HSS \mid \sim SQN_{IMSI}} \\
\displaystyle &\frac{UE \mid \equiv \#SQN_{IMSI}, UE \mid \equiv HSS \mid \sim SQN_{IMSI}} {UE \mid \equiv HSS \mid \equiv SQN_{IMSI}}\\
\displaystyle &\frac{UE \mid \equiv HSS \mid \Rightarrow SQN_{IMSI}, UE \mid \equiv HSS \mid \equiv SQN_{IMSI}} {UE\mid \equiv SQN_{IMSI}}  \label{ue_imsi}
\end{split}
\end{equation} 
\noindent 3) When the HSS gets $ RES $ from the UE, where $ RES $ is defined by Eq.~(\ref{res}), the HSS knows that the authentication is completed.  According to the message-meaning rule in Eq.~(\ref{mmr}) and the nonce-verification rule in Eq.~(\ref{nvr}), we get:
 \begin{equation}
 \begin{split}
 RES = f_2( Rand, Key )   \label{res}
 \end{split}
 \end{equation}
\begin{equation}
\begin{split}
\displaystyle &\frac{HSS \mid \equiv UE \stackrel{Key}{\longleftrightarrow} HSS, HSS \vartriangleleft [RAND]_{Key}} {HSS \mid \equiv UE \mid \sim RAND} \\
\displaystyle &\frac{HSS \mid \equiv \#(RAND), HSS \mid \equiv UE \mid \sim RAND} {HSS \mid \equiv UE \mid \equiv RAND} \label{rand}
\end{split}
\end{equation} 
In this process, because the UE needs to check $SQN_{IMSI}$ and then sends $ RES $ to the HSS, the HSS believes that the UE recognizes $ RAND $, only the UE believes $SQN_{IMSI}$ first. Taking into account the dependency relationship between $ RAND $ and $SQN_{IMSI}$, we add a dependency relationship rule, shown in Eq.~(\ref{dr}):
\begin{equation}
\begin{split}
\frac{X \rightarrow Z, P\mid \equiv Q \mid \equiv X} {P\mid \equiv Q \mid \equiv Z} \label{dr}
\end{split}
\end{equation}
where X $\rightarrow$ Z means that X depends on Z. From Eq.~(\ref{rand}) and Eq.~(\ref{dr}), we get:
\begin{equation}
\begin{split}
\frac{RAND \rightarrow SQN_{IMSI}, HSS \mid \equiv UE \mid \equiv RAND} {HSS\mid \equiv UE \mid \equiv SQN_{IMSI}} \label{SQNIMSI}
\end{split}
\end{equation}
\noindent 4) For next authentications, the UE must has an agreement on the shared pseudonym $P_i$ with the HSS. Here we give the proof of the goal 3 and 4. According to Eq.~(\ref{ue_imsi}), Eq.~(\ref{SQNIMSI}) and the session key rule in Eq.~(\ref{sr}), we get:
\begin{equation}
\begin{split}
\displaystyle &\frac{UE \mid \equiv \# (SQN_{IMSI}), UE \mid \equiv HSS \mid \equiv SQN_{IMSI}} {UE \mid \equiv UE \stackrel{SQN_{IMSI}}{\longleftrightarrow} HSS}\\ 
\displaystyle &\frac{HSS \mid \equiv \# (SQN_{IMSI}), HSS \mid \equiv UE \mid \equiv SQN_{IMSI}} {HSS \mid \equiv UE \stackrel{SQN_{IMSI}}{\longleftrightarrow} HSS}\\ \label{ue}
\end{split}
\end{equation}
When the UE uses $P_i$ as his pseudonym, $P_i$ can be regarded as an encrypted IMSI which is protected by $SQN_{IMSI}$. According to Eq.~(\ref{ue}), the message-meaning rule in Eq.~(\ref{mmr}), the nonce-verification rule in Eq.~(\ref{nvr}) and the jurisdiction rule in Eq.~(\ref{jr}), we get:
\begin{equation}
\begin{split}
\displaystyle &\frac{HSS \mid \equiv UE \stackrel{(Key,SQN_{IMSI})}{\longleftrightarrow} HSS, HSS \vartriangleleft [IMSI]_{(Key,SQN_{IMSI})}} {HSS \mid \equiv UE \mid \sim [IMSI]_{SQN_{IMSI}}}\\ 
\displaystyle &\frac{HSS \mid \equiv \# ([IMSI]_{SQN_{IMSI}}), HSS \mid \equiv UE \mid \sim [IMSI]_{SQN_{IMSI}}} {HSS \mid \equiv UE \mid \equiv [IMSI]_{SQN_{IMSI}}}\\
\displaystyle &\frac{HSS \mid \equiv UE \mid \Rightarrow [IMSI]_{SQN_{IMSI}}, HSS \mid \equiv UE \mid \equiv [IMSI]_{SQN_{IMSI}}} {HSS \mid \equiv [IMSI]_{SQN_{IMSI}}}
\end{split}
\end{equation}
In the normal authentication, we get: HSS $\mid \equiv P_i$ and UE $\mid \equiv SQN_{IMSI}$, so the HSS can authenticate the UE's IMSI and update $P_i$ with the UE synchronously.
\subsection{The Anonymity of the UE's Identity}
In our scheme, the shared pseudonym is adopted to resist the intersection attack and the mark attack. With the variable k-pseudonym sets, the adversary cannot identify the shared pseudonym without the shared key. In the worst case, we use \{IMSI\} in the initial access authentication, which has the same anonymity with the basic k-pseudonym scheme. But we emphasize that the \{IMSI\} only appears once in our system and the adversary cannot recognize the initial authentication easily. In next authentications, the UE utilize the variable shared pseudonym as his temporary identity, so our scheme performs better in general. Even when the unsynchronised pseudonym comes, the UE can continue the access authentication with the $P_0$. Because the $Key$ and the $SQN_{IMSI}$ (or $SQN_{P_0}$) are the stable shared information between the UE and the HSS, it is a reasonable assumption that the UE and the HSS can get $P_0$ synchronously. Next, we analyze brute-force attack. In the scheme, we initialize ZUC with the $Key$ and the $SQN_{IMSI}$ (or $SQN_{P_0}$), and only the UE and the HSS can generate a series of relevant shared pseudonyms legally. Under such conditions, the probability of getting the UE's IMSI (especially MSIN) by the exhaustive method is $\frac{1}{2^{40}}$. But if the attacker want trace or mark the UE, he also has to guess the shared key, which is the vital part to get the UE's next shared psedudonym. This means the probability of recognizing UE's current valid identity is $\frac{1}{2^{168}}$, where the $Key$ is 128 bits and MSIN is 40 bits.

\begin{comment}
 , an ability to meet the complicated network environments.
 
\end{comment}

\section{Conclusions}
In this paper, we propose a shared key based anonymous authentication scheme in 5G access authentication. And by the shared pseudonym, the UE can construct the variable k-pseudonym sets in subsequent access authentications. Moreover, owing to the variable k-pseudonym sets, our scheme can resist the intersection attack and the mark attack. We also give the shared pseudonym construction method and the recovery mechanism for the asynchronous situations. Finally, after BAN logic analysis, we conclude that the UE and the HSS can get the variable shared pseudonym ($P_i$) synchronous and privately. Besides, we hope to find some ways to improve the communication efficiency and reduce the communication latency.

\begin{comment}
In this paper, we propose a shared key based anonymous authentication scheme in 5G access authentication. And by the shared pseudonym, our scheme can resist the intersection attack and the mark attack. In the subsequent access authentications, the UE can use the variable shared pseudonym as his temporary identity. Meanwhile, we give the shared pseudonym construction method and the recovery mechanism for the unsynchronised situations. Finally, after BAN logic analysis, we conclude that the UE and the HSS can get the variable shared pseudonym ($P_i$) synchronous and privately. Besides, we hope to find some ways to improve the communication efficiency and reduce the communication latency.

\end{comment}

% ---- Bibliography ----
%
% BibTeX users should specify bibliography style 'splncs04'.
% References will then be sorted and formatted in the correct style.
%
% \bibliographystyle{splncs04}
% \bibliography{mybibliography}
%
\bibliographystyle{splncs04}
\bibliography{ccs-sample}

\begin{thebibliography}{10}
\providecommand{\url}[1]{\texttt{#1}}
\providecommand{\urlprefix}{URL }
\providecommand{\doi}[1]{https://doi.org/#1}

\bibitem{S3161379}
3GPP: {3GPP Discussion Document S3-161379 and S3-161380}.
  \url{http://www.3gpp.org/ftp/tsg_sa/WG3_Security/TSGS3_84b_San_Diego/Docs/}
  (2016), last accessed on 25-11-2018.

\bibitem{S3162108}
3GPP: {3GPP Discussion Document S3-162108}.
  \url{http://www.3gpp.org/ftp/tsg_sa/WG3_Security/TSGS3_84b_San_Diego/Docs/}
  (2016), last accessed on 25-11-2018.

\bibitem{ahlawat2018investigating}
Ahlawat, A., Kumar, S.: Investigating various possible attacks and
  vulnerabilties in lte  (2018)

\bibitem{av}
Al-Saraireh, J., Yousef, S.: A new authentication protocol for umts mobile
  networks. Eurasip Journal on Wireless Communications \& Networking
  \textbf{2006}(1),  098107 (2006)

\bibitem{alliance20155g}
Alliance, N.: 5g white paper. Next generation mobile networks, white paper pp.
  1--125 (2015)

\bibitem{cerwall2016ericsson}
Cerwall, P.: Ericsson mobility report.(june 2017). White Paper  (2016)

\bibitem{MNET}
Chaouchi, H., Laurent-Maknavicius, M.: Wireless and mobile network security.
  John Wiley \& Sons (2013)

\bibitem{analy}
Cobo~Jim{\'e}nez, E.: Encrypting imsi to improve privacy in 5g networks (2017)

\bibitem{PKI}
Ekene, O.E., Ruhl, R., Zavarsky, P.: Enhanced user security and privacy
  protection in 4g lte network. In: Computer Software and Applications
  Conference (COMPSAC), 2016 IEEE 40th Annual. vol.~2, pp. 443--448. IEEE
  (2016)

\bibitem{ferrag2017security}
Ferrag, M.A., Maglaras, L., Argyriou, A., Kosmanos, D., Janicke, H.: Security
  for 4g and 5g cellular networks: A survey of existing authentication and
  privacy-preserving schemes. Journal of Network and Computer Applications
  (2017)

\bibitem{lte2}
Forsberg, D., Horn, G., Moeller, W.D., Niemi, V.: LTE security. John Wiley \&
  Sons (2012)

\bibitem{goyal2006attribute}
Goyal, V., Pandey, O., Sahai, A., Waters, B.: Attribute-based encryption for
  fine-grained access control of encrypted data. In: Proceedings of the 13th
  ACM conference on Computer and communications security. pp. 89--98. Acm
  (2006)

\bibitem{GUTI}
Hong, B., Bae, S., Kim, Y.: Guti reallocation demystified: Cellular location
  tracking with changing temporary identifier. In: Symposium on Network and
  Distributed System Security (NDSS). ISOC (2018)

\bibitem{sqn}
Khan, M.S.A., Mitchell, C.J.: Improving air interface user privacy in mobile
  telephony. In: International Conference on Security Standardisation Research.
  pp. 165--184 (2015)

\bibitem{khan2018imsi}
Khan, M., Ginzboorg, P., Niemi, V.: Imsi-based routing and identity privacy in
  5g. In: Proceedings of the 22nd Conference of Open Innovations Association
  FRUCT, Jyvaskyla, Finland (2018)

\bibitem{IBE}
Khan, M., Niemi, V.: Concealing imsi in 5g network using identity based
  encryption. In: Yan, Z., Molva, R., Mazurczyk, W., Kantola, R. (eds.) Network
  and System Security. pp. 544--554. Springer International Publishing, Cham
  (2017)

\bibitem{kim2013design}
Kim, S.: A design of MILENAGE algorithm-based mutual authentication protocol
  for the protection of initial identifier in LTE. Ph.D. thesis, Master’s
  thesis, Soongsil University Google Scholar (2013)

\bibitem{gsm}
Lee, C.C., Hwang, M.S., Yang, W.P.: Extension of authentication protocol for
  gsm. IEE Proceedings-Communications  \textbf{150}(2),  91--95 (2003)

\bibitem{lte1}
Lee, M.F., Smart, N.P., Warinschi, B., Watson, G.J.: Anonymity guarantees of
  the umts/lte authentication and connection protocol. International journal of
  information security  \textbf{13}(6),  513--527 (2014)

\bibitem{kli}
Li, X., Liu, H., Wei, F., Ma, J., Yang, W.: A lightweight anonymous
  authentication protocol using k-pseudonym set in wireless networks. In:
  Global Communications Conference (GLOBECOM), 2015 IEEE. pp.~1--6. IEEE (2015)

\bibitem{PSE}
Norrman, K., N{\"a}slund, M., Dubrova, E.: Protecting imsi and user privacy in
  5g networks. In: Proceedings of the 9th EAI International Conference on
  Mobile Multimedia Communications. pp. 159--166. ICST (Institute for Computer
  Sciences, Social-Informatics and Telecommunications Engineering) (2016)

\bibitem{zuc2}
Orhanou, G., El~Hajji, S., Lakbabi, A., Bentaleb, Y.: Analytical evaluation of
  the stream cipher zuc. In: Multimedia Computing and Systems (ICMCS), 2012
  International Conference on. pp. 927--930. IEEE (2012)

\bibitem{rupprecht2018security}
Rupprecht, D., Dabrowski, A., Holz, T., Weippl, E., P{\"o}pper, C.: On security
  research towards future mobile network generations. IEEE Communications
  Surveys \& Tutorials  (2018)

\bibitem{3gpp3g}
SAGE, E.: Specification of the milenage algorithm set: an example algorithm set
  for the 3gpp authentication and key generation functions f1, f1*, f2, f3, f4,
  f5 and f5*

\bibitem{BAN1}
Wen, J., Zhang, M., Li, X.: The study on the application of ban logic in formal
  analysis of authentication protocols. In: Proceedings of the 7th
  international conference on Electronic commerce. pp. 744--747. ACM (2005)

\bibitem{zuc1}
Wu, H., Nguyen, P.H., Wang, H., Ling, S.: Cryptanalysis of the stream cipher
  zuc in the 3gpp confidentiality \& integrity algorithms 128-eea3 \& 128-eia3.
  Rump session of Asiacrypt  \textbf{2010} (2010)

\bibitem{sa}
Zhang, M., Fang, Y.: Security analysis and enhancements of 3gpp authentication
  and key agreement protocol. IEEE Transactions on wireless communications
  \textbf{4}(2),  734--742 (2005)

\end{thebibliography}
\end{document}